Is non-informative Bayesian analysis appropriate for wildlife management: survival of San Joaquin Kit Fox and declines in amphibian populations

Subhash R. Lele

Department of Mathematical and Statistical Sciences

University of Alberta

Edmonton, AB T6G 2G1

Canada

Email: slele@ualberta.ca

Corresponding author: Subhash R Lele, Department of Mathematical and Statistical Sciences, University of Alberta, Edmonton, AB T6G 2G1 Canada.

Email: slele@ualberta.ca, Phone: 780 492 4290 Fax: 780 492 6826

*Statement of authorship*: SRL conceived of the ideas, wrote the manuscript and did the statistical programming.

**Abstract:**

Computational convenience has led to widespread use of Bayesian inference with vague or flat priors to analyze state-space models in ecology. Vague priors are claimed to be objective and to let the data speak. Neither of these claims is valid. Statisticians have criticized the use of vague priors from philosophical to computational to pragmatic reasons. Ecologists, however, dismiss such criticisms as empty philosophical wonderings with no practical implications. We illustrate that use of vague priors in population viability analysis and occupancy models can have significant impact on the analysis and can lead to strikingly different managerial decisions. Given the wide spread applicability of the hierarchical models and uncritical use of non-informative Bayesian analysis in ecology, researchers should be cautious about using the vague priors as a default choice in practical situations.

**Introduction**

Hierarchical models, also known as state-space models, mixed effects models or mixture models, have proved to be extremely useful for modeling and analyzing ecological data (e.g. Kery and Schaub 2012, Bolker 2008). Although these models can be analyzed using the likelihood methods (Lele et al. 2007, Lele et al. 2010), the Bayesian approach is the most advocated approach for such models. Many researchers even name hierarchical models as 'Bayesian models' (Parent and Rivot 2013). Of course, there are no Bayesian models or frequentist models. There are only statistical models that we fit to the data using either a Bayesian approach or a frequentist approach. The subjectivity of the Bayesian approach is bothersome to most scientists (Efron 1986; Dennis 1996) and hence the trend is to use the non-informative, also called vague or objective, priors instead of the subjective priors provided by the expert. These non-informative priors purportedly "let the data speak" and do not bias the conclusions with the subjectivity inherent in the subjective priors. It has been claimed that Bayesian inferences based on non-informative priors are similar to the likelihood inference (e.g. Clark, 2005, pages 3 and 5) although such a result has never been rigorously established. The fact is that it is not even clear what a non-informative prior really means. There are many different ways to construct non-informative priors (Press 2002, Chapter 5). The most commonly used non-informative priors are either the uniform priors or the priors with very large variances spreading the probability mass almost uniformly over the entire parameter space. These priors have been criticized on computational grounds (e.g. Natarajan and McCulloch, 1998) because they can inadvertently lead to non-sensible posterior distributions. More fundamentally, one of the founders of modern statistics, R.A. Fisher, objected to the use of flat priors because of their lack of invariance under transformation (deValpine 2009; Lele and Dennis 2009). For example, a uniform prior on (0,1) for the probability of success in a Binomial model turns into a non-uniform prior on the logit scale (See figure 1a, 1b). If a uniform prior is supposed to express complete ignorance about different parameter values, then this says that if one is ignorant about $p$, one is quite informative about $\log \frac{p}{1-p}$. Similarly a normal prior with large variance on the logit scale, that presumably represents complete ignorance, transforms into a non-uniform prior on the probability scale (see figure 1c, 1d).

----------------------------------------- Figure 1 here ----------------------------------------

This makes no sense because they are one-one transformations of each other; if we are ignorant about one, we should be equally ignorant about the other. Press (2002, Chapter 5) provides an excellent review of various problems associated with the definitions and use of non-informative priors along with interesting historical notes. Unfortunately, ecologists and practitioners tend to dismiss these criticisms; considering them as empty philosophical wonderings of statisticians with no practical relevance (e.g. Clark 2005). The goal of this paper is to disabuse the ecologists of the notion that there is no difference between non-informative Bayesian inference and likelihood-based inference and that the philosophical underpinnings of statistical inference are irrelevant to practice. To illustrate this point, we consider two important ecological problems: Population monitoring and population viability analysis. We show that, due to lack of invariance, analysis of the same data under the same statistical model can lead to substantially different conclusions under the non-informative Bayesian framework. This is disturbing because common sense dictates that same data and same model should lead to the same scientific conclusions. The problem with the non-informative priors is that they do not 'let the data speak'; contrary to what is commonly claimed, they bring in their own biases in the analysis. The goal of this paper is not to suggest that the likelihood analysis, which is generally parameterization invariant, is the only way or the right way to do the data analysis in applied ecology. That debate is subtle, potentially unresolvable and is best left for another place and time. The only goal of this paper is to show the practical implications of the lack of invariance of the non-informative priors that we feel are significant for wildlife managers.

**Population viability analysis (PVA) under the Ricker model:**

Let us consider the San Joaquin Kit Fox data set used by Dennis and Otten (2000) who originally analyzed these data. This kit fox population inhabits a study area of size 135 km$^2$ on the Navel Petroleum Reserves in California (NPRC). The abundance time-series for the years 1983-1995 was obtained to conduct an extensive population dynamics study as part of the NPRC Endangered Species and Cultural Resources Program. The annual abundance estimates were obtained from capture-recapture histories generated by trapping adult and yearling foxes each winter between 1983-1995. We refer the reader to Dennis and Otten (2000) for further details on these data and abundance estimation technique.

Dennis and Otten (2000) analyzed these data using the Ricker model. The deterministic version of the Ricker model can be written in two different but mathematically equivalent forms. It may be written in terms of the growth parameter $a$ and density dependence parameter $b$ as $\log N_{t+1} - \log N_t = a + bN_t$ or in terms of growth parameter $a$ and carrying capacity parameter $K$ as $\log N_{t+1} - \log N_t = a\left(1 - \frac{N_t}{K}\right)$. It is reasonable to expect that the conclusions about the survival of the San Joaquin Kit Fox population would remain the same whether one uses the $(a,b)$ formulation or the $(a,K)$ formulation. In statistical jargon, we call this change in the form of the model reparameterization and we will use this term, instead of the term different formulation, in the rest of the paper. Following Dennis and Otten (2000), we use a stochastic version of the Ricker model where the parameter $a$, instead of being fixed, varies randomly from year to year. The abundance values are themselves an estimate of the true abundances and hence we consider the sampling variability in the model as well. The standard errors for the abundance estimates were nearly proportional to the abundance estimates and hence the Poisson sampling distribution makes reasonable sense. The full model can be written as a state-space model as follows. Let $X_t = \log N_t$.

We call the following form of the model the $(a,b)$ parameterization.

a) Process model: $X_{t+1} | X_t \sim N(X_t + a + b\exp(X_t), \sigma^2)$ where $b$ is the density dependence parameter.

b) Observation model: $Y_t | X_t \sim Poisson(\exp(X_t))$

One can write this model in an alternative form that we call the $(a,K)$ parameterization.

a) Process model: $X_{t+1} | X_t \sim N(X_t + a\left(1 - \frac{\exp(X_t)}{K}\right), \sigma^2)$ where $K$ is the carrying capacity.

b) Observation model: $Y_t | X_t \sim Poisson(\exp(X_t))$

These two models are mathematically identical to each other. Our goal is to fit these models to the observed data and conduct population viability analysis using the population prediction intervals (PPI) (Saether et al. 2000). Common sense dictates that because the data are the same and the models are mathematically equivalent to each other, the PPI computed under the two parameterizations should also be identical to each other.

We use Bayesian inference using non-informative priors to compute PPI under these two forms. For the Bayesian inference, we need to specify the priors on the parameters. We use the following non-informative priors for the parameters in the respective parameterization.

Priors for the $(a,b)$ parameterization: $a \sim N(0,10)$, $b \sim U(0,1)$, $\sigma^2 \sim LN(0,10)$

Priors for the $(a,K)$ parameterization

$a \sim N(0,10)$, $K \sim Gamma(100,100)$, $\sigma^2 \sim LN(0,10)$

For comparison, we use data cloning algorithm (Lele et al. 2007, 2010) to compute the likelihood-based PPI under these two parameterizations. The analysis was conducted using the package 'dclone' (Solymos, 2010) in the R software. The parameter estimates are given in the table below.

------------------------------------------- Table 1 here -------------------------------------------

Notice that the parameter estimates for the two parameterizations are quite a bit different; on the other hand, the maximum likelihood estimates (MLE) under two parameterizations are nearly identical to each other under both parameterizations as they should be. The small differences are due to the Monte Carlo error.

In figure 2 we show the PPI obtained under the likelihood and the non-informative Bayesian approach.

------------------------------------------- Figure 2 here -------------------------------------------

One can make two important observations: (1) The PPI obtained under the $(a,b)$ parameterization and the PPI obtained under the $(a,K)$ parameterization, both obtained under purportedly non-informative priors, are quite different. Depending on which parameterization the researcher happens to use, the scientific conclusions will be quite different. This, if not totally unacceptable, is at least disturbing. As we said earlier, same data, same model should lead to the same conclusions. However, non-informative Bayesian analysis does not satisfy this common sense requirement. (2) The likelihood based PPI is quite different than the non-informative prior based PPI. Contrary to what is commonly claimed, the non-informative priors do not lead to inferences that are similar to the likelihood inferences.

**Occupancy models and the decline of amphibians:**

One of the central tasks an applied ecologist is entrusted with is to monitor the existing populations. These monitoring data are the input to many further ecological analyses. We consider the following simple model that is commonly used in analyzing occupancy data with replicate visits (MacKenzie et al. 2002). We denote probability of occupancy by $\psi$ and probability of detection by $p$. For simplicity (and, to emphasize that these results do not happen only for complex models), we assume these do not depend on covariates. We assume there are $n$ sites and each site is visited $k$ times. Other assumptions about close population and independence of the surveys are similar to the ones described in MacKenzie et al. (2002). The replicate visit model can be written as follows.

Hierarchy 1: $Y_i \sim Bernoulli(\psi)$ for $i = 1,2,...,n$

Hierarchy 2: $O_{ij} | Y_i = 1 \sim Bernoulli(p)$ where $j = 1,2,..,k$

We assume that if $Y_i = 0$, then $O_{ij} = 0$ with probability 1 for $j = 1,2,..,k$. That is, there are no false detections. This model can be written in terms of logit parameters as follows:

Hierarchy 1: $Y_i \sim Bernoulli(\beta)$ for $i = 1,2,...,n$ where $\beta = \log\frac{\psi}{1-\psi}$

Hierarchy 2: $O_{ij} | Y_i = 1 \sim Bernoulli(\delta)$ where $j = 1,2,..,k$ where $\delta = \log\frac{p}{1-p}$

The second parameterization is commonly used when there are covariates and the logit link is used to model the dependence of the covariates on the occupancy and detection probabilities. We use the following non-informative priors for the two parameterizations.

The $(\psi, p)$ parameterization: $\psi \sim uniform(0,1)$, $p \sim uniform(0,1)$

The $(\theta, \delta)$ model: $\theta \sim N(0,1000)$, $\delta \sim N(0,1000)$

These are commonly used non-informative priors on the respective scales. The goal of the analysis is to predict the total occupancy rate. To compute this, we need to compute the probability that a site that is observed to be unoccupied is, in fact, occupied. We need to compute $P(Y_i = 1 | O_{ij} = 0, j = 1,2,...,k)$. We can compute it by using standard conditional probability arguments as:

$$P(Y_i = 1 | O_{ij} = 0, j = 1,2,...,k) = \frac{P(O_{ij} = 0, j = 1,2,...,k | Y_i = 1)P(Y_i = 1)}{P(O_{ij} = 0, j = 1,2,...,k)}$$

$$= \frac{(1-p)^k \psi}{(1-p)^k \psi + (1-\psi)}$$

We first present a simulation study where we show the differences in the non-informative Bayesian inferences between the two parameterizations. We present the simulation results for the case of 30 sites and two visits to each site. We consider three different combinations of probability of detection and probability of occupancy; both small, occupancy large but detection small and occupancy small and detection large. It is well known (e.g. Walker, 1969) that as the sample size increases, Bayesian inferences become similar to the likelihood inference. We checked our program (provided in the supplementary information) by taking 100 sites and 20 visits per site. For these sample sizes, as expected, the inferences were nearly invariant.

----------------------------------------- Table 2 here -----------------------------------------

Table 2 shows that the inferences about point estimates of the probability of occupancy and detection and more importantly about the probability that a site is, in fact, occupied when it is observed to be unoccupied on both visits are not invariant to the parameterization. This has significant practical implications: The predicted occupancy rates will be quite different depending on which parameterization is used.

How does this work out in real life situation? Let us reanalyze the data presented in MacKenzie et al. (2002). We consider a subset of the occupancy data for American Toad (*Bufo Americanas*) where we only consider the first three visits. There are 27 sites that have at least three visits. The raw occupancy rate, the proportion of sites occupied at least once in three visits, was 0.37. We fit the constant occupancy and constant detection probability model using the two different parameterizations described above. The point estimates of various quantities are shown in the table below.

----------------------------------------- Table 3 here -----------------------------------------

The differences in the two analyses are striking. According to one analysis, we will declare an unoccupied site to have probability of being occupied as 0.296 where as the other analysis will replace a 0 by 1 with probability 0.6715, more than double the first analysis. Given the data, after adjusting for detection error, we will declare the study area to have occupancy rate to be 0.56 under one analysis but under the other analysis, we will declare it to be 0.80. In figure 3, we show the posterior distributions for the occupancy rate under the two parameterizations. It is obvious that the decisions based on these two posterior distributions are likely to be very different.

Now imagine facing a lawyer in the court of law or a politician who is challenging the results of the wildlife manager who is testifying that the occupancy rates are too low (or, too high for invasive species). All they have to do, while still claiming to do a legitimate non-informative analysis, is use a parameterization that gives different results to raise the doubt in the minds of the jurors or the senators on the committee. This is not a desirable situation.

**Discussion:**

Using different parameterizations of a statistical model depending on the purpose of the analysis is not uncommon. For example, in survival analysis the exponential distribution is written using the hazard function or the mean survival function depending on the goal of the study. They are simply reciprocals of each other. Similarly Gamma distribution is often written in terms of rate and shape parameter or in terms of mean and variance that is suitable for regression models. Beta regression is presented in two different forms: Regression models for the two shape parameters or regression model for the mean keeping variance parameter constant (Ferrari and Cribari-Neto, 2004). All these situations present a problem for flat and other non-informative priors because same data and same model can lead to different conclusions depending on which parameterization is used. The issue of choice of default priors and its impact on statistical inference has also arisen in genomics (Rannala et al. 2013). One can possibly construct similar examples in the Mark-Capture-Recapture methods where different parameterizations are commonly used. The examples presented in this paper are likely to be more a rule than exceptions.

The lack of parameterization invariance of the flat priors is a long known criticism. This criticism was potent enough that it needed addressing. Harold Jeffreys tried to construct priors that yield parameterization invariant conclusions. They are now known as Jeffreys priors. A full description of these priors and how to construct them is beyond the scope of this paper (See Press 2002 for easily accessible details). However, it suffices to say that they are proportional to the inverse of the expected Fisher information matrix. In order to construct them, one needs to know the likelihood function and the exact analytic expression for the expected Fisher information matrix. This is seldom available for hierarchical models. But simply to illustrate these priors, consider a simple example where $Y | p \sim Binomial(N, p)$. The Jeffrey's prior for the probability of success is $Beta(0.5, 0.5)$ and is plotted in Figure 4.

------------------------------------------------- Figure 4 here --------------------------------------------

Even a quick look at this figure will convince the reader that the prior is nowhere close to looking like what one would consider a non-informative prior. It is highly concentrated near 0 and 1 with very small weight in the middle. Even when Jeffreys prior can be computed, it will be difficult to sell this prior as an objective prior to the jurors or the senators on the committee. The construction of Jeffreys and other objective priors for multi-parameter models poses substantial mathematical difficulties. The common practice is to put independent priors on each of the parameter. Why such prior knowledge of independence of the parameters be considered 'non-informative' is completely unclear. It seems to be more a matter of convenience than a matter of principle.

To summarize, we have shown that non-informative priors neither 'let the data speak' nor do they correspond (even roughly) to likelihood analysis. They seem to add their own biases in the scientific conclusions. Just because the euphemistic terms such as objective priors, non-informative priors or objective Bayesian analysis are used, it does not mean that the analyses are not subjective. A truly subjective prior based on expert opinion is, perhaps, preferable to the non-informative priors because in the former case the subjectivity is clear and well quantified (and, may be justified) whereas in the latter the subjectivity is hidden and not quantified. Many applied ecologists are using the non-informative Bayesian approach almost as a panacea to deal with hierarchical models believing that they are presenting objective, unbiased results. The resultant analysis, because of the lack of invariance to parameterization, has unstated and unquantifiable biases and hence may not be justifiable for either the scientific purposes or the managerial applications.

**Acknowledgement:**

This work was carried out while the author was visiting the Center for Biodiversity Dynamics, NTNU, Norway and the Department of Biology, Sun Yat Sen University, Guangzhou. This work was supported by a grant from National Science Engineering Research Council of Canada. H. Beyer, C. McCulloch, J. Ponciano, P. Solymos provided useful comments.

Table 1: Parameter estimates for the Kit Fox data using different parameterizations and non-informative priors and maximum likelihood.

| Parameter | Bayes (a,b) | Bayes (a,K) | MLE (a,b) | MLE (a,K) |
|---|---|---|---|---|
| $a$ | 0.7542 | 0.4812 | 0.7404 | 0.7322 |
| $K$ | 159.6425 | 141.39 | 160.1643 | 159.7164 |
| $\sigma$ | 0.4916 | 0.5053 | 0.4360 | 0.4358 |

Table 2: Simulation study showing the effect of using different parameterizations on the Bayesian estimation of occupancy and detection parameters using non-informative priors. Total number of sites is 30 and each site is visited 2 times.

| Parameter | $p = 0.3, \psi = 0.3$ | | $p = 0.8, \psi = 0.3$ | | $p = 0.3, \psi = 0.8$ | |
|---|---|---|---|---|---|---|
| | Probability | Logit | Probability | Logit | Probability | Logit |
| $\hat{p}$ | 0.3079 | 0.1864 | 0.7394 | 0.7855 | 0.3648 | 0.2950 |
| $\hat{\psi}$ | 0.4168 | 0.7786 | 0.3438 | 0.3240 | 0.6904 | 0.9174 |
| $P(Y=1|O=0)$ | 0.2535 | 0.7054 | 0.03243 | 0.0196 | 0.4581 | 0.8567 |

Table 3: Parameter estimates for the American Toad occupancy data using non-informative Bayesian analysis under different parameterization

| Parameter | Bayes probability | Bayes Logit |
|---|---|---|
| $p$ | 0.3245 | 0.2314 |
| $\psi$ | 0.5770 | 0.8183 |
| $P(Y=1|O=0)$ | 0.2960 | 0.6715 |
| Total occupancy rate | 0.5568 | 0.7932 |

**Figure legends:**

Figure 1: Non-informative prior on one scale is informative on a different scale. What is considered non-informative on the logit scale will be considered quite informative on the probability scale and what is considered non-informative on the probability scale will be considered informative on the logit scale.

Figure 2: Population Prediction Intervals (PPI) for the Kit Fox data using non-informative Bayesian analysis under two different parameterizations and the maximum likelihood analysis. Notice that non-informative Bayesian analysis does not approximate the maximum likelihood analysis and depends on the specific parameterization.

Figure 3: Jeffreys non-informative prior, which has invariance property, on the probability scale is concentrated near 0 and 1 with very little weight for the in the middle.

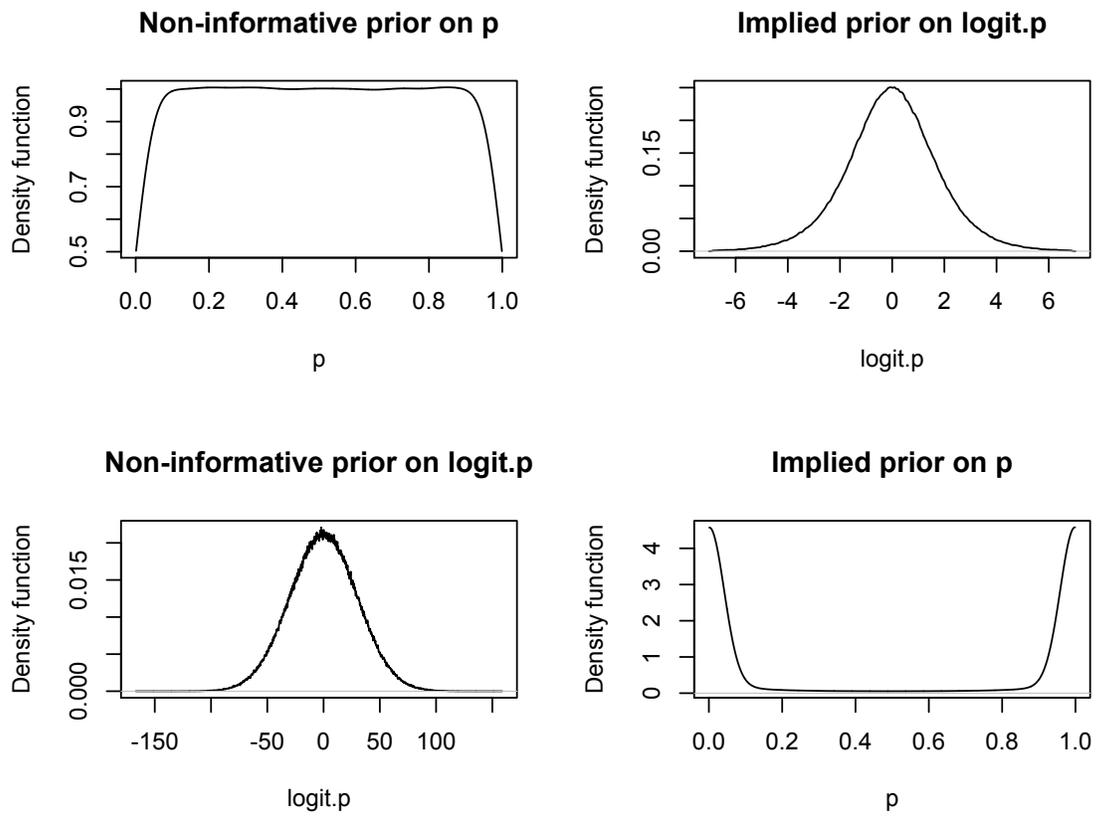

**Figure 1**

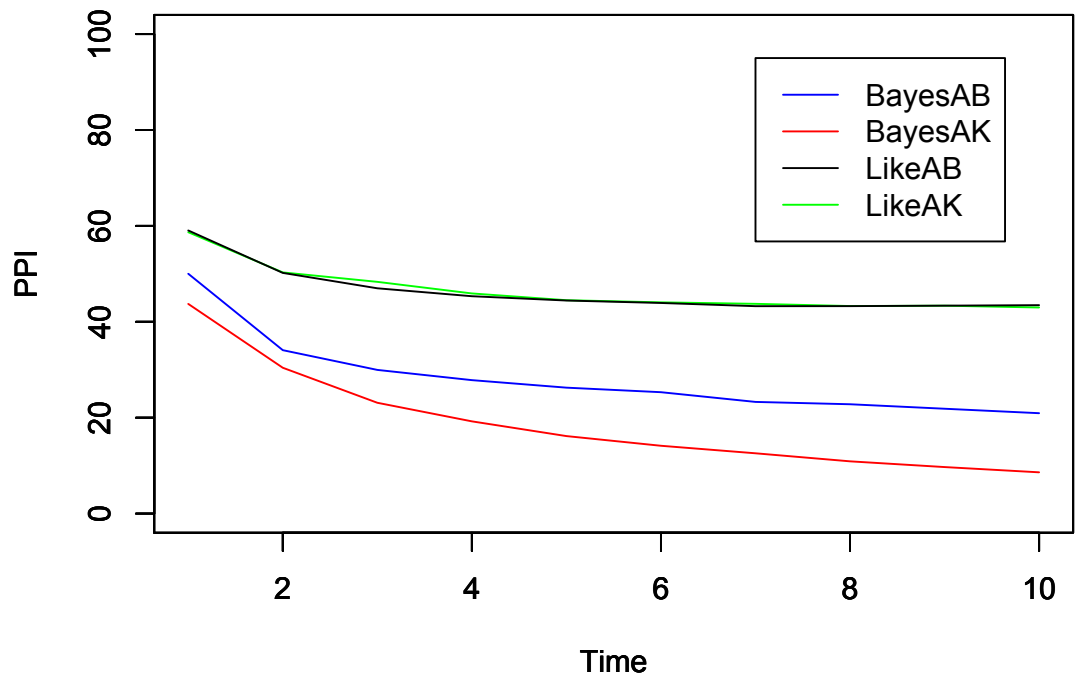

Figure 2

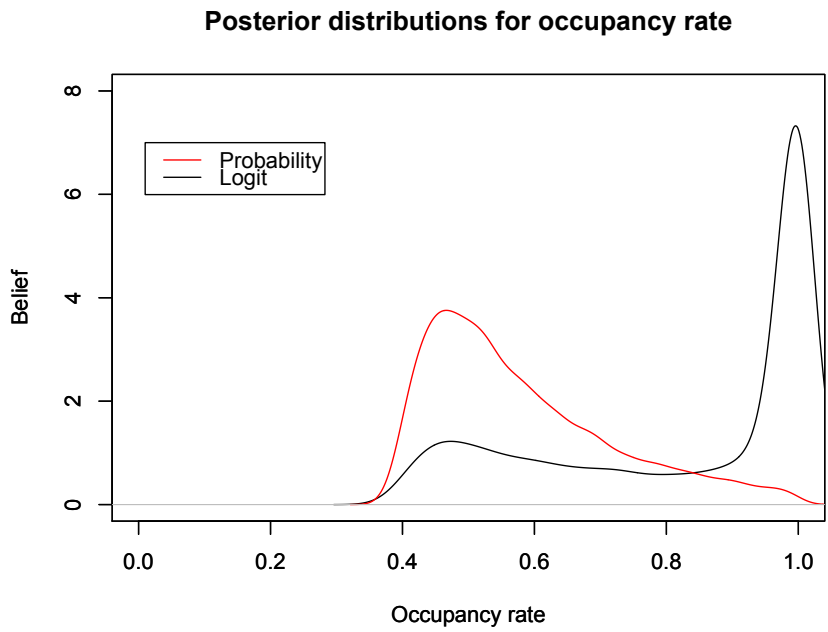

**Figure 3**

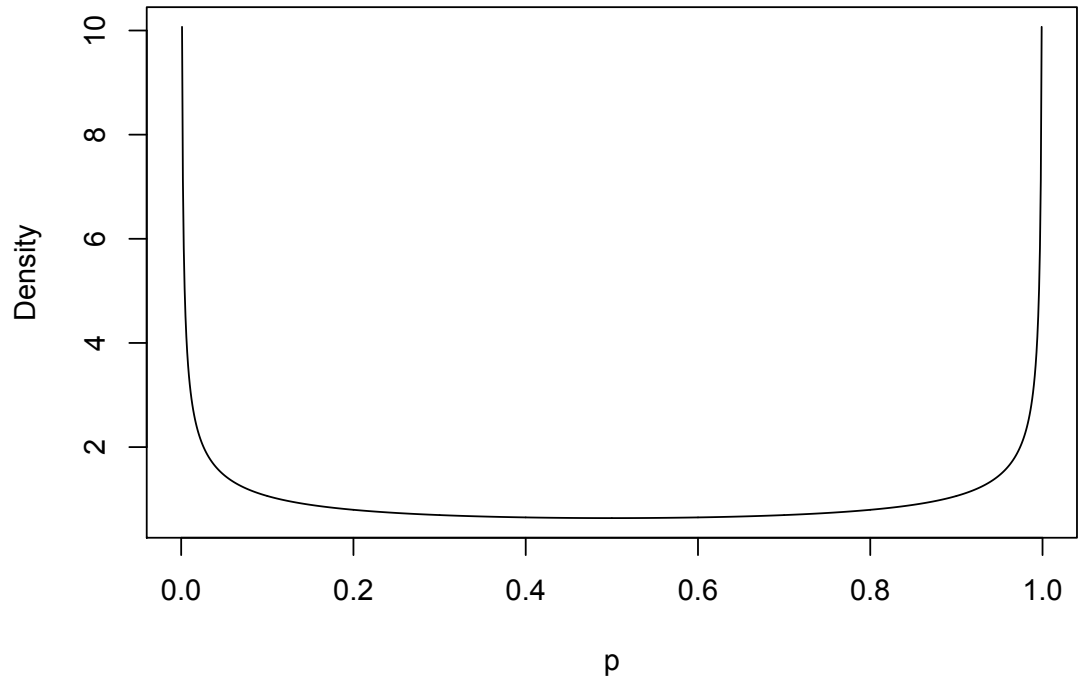

**Figure 4**